\documentstyle[12pt,epsf,a4]{article}

\begin{document}

\title{Unusual valence, negative charge-transfer gaps and self-doping
in transition-metal compounds\footnote{This paper is devoted to the
memory of Prof.~R.~Dagys and was originally published in a memorial
volume of Lithuanian Journal of Physics (Lietuvos fizikos \v
zurnalas), {\bf 37}, 65 (1997)}}

\author{D.~Khomskii\\
\\Laboratory of Solid State Physics,
\\Groningen University,
\\Ni\kern-1pt jenborgh 4, 9747 AG Groningen,
\\The Netherlands}

\date{}

\maketitle

\begin{abstract}
In this paper I discuss the electronic structure
and properties of a specific, rather unconventional class of
transition metal (TM) compounds, e.g.\ TM oxides, which
formally have unusually high values of the oxidation state, or
valence, of TM\null.  In contrast to the typical situation, in this
case the charge-transfer gap (excitation energy for the transfer of
electrons from the ligands to the TM) is very small and may even
become negative. As a result a profound modification of an electronic
structure and of all the properties may take place: there appear holes
in the oxygen $p$-bands (``self-doping''); the material may become the
metal of a specific type; there may occur insulator--metal transitions
of a specific type; magnetic properties may be quite different from
the ones expected normally; the character of elementary excitations
may change drastically.  I give general discussion of such situation
and consider several examples of corresponding systems and their
specific properties.
\end{abstract}

\section{Introduction}
In this paper the specific class of transition metal compounds will be
discussed\hskip0pt---the systems with the negative charge transfer
gap.  This article is neither a fully original paper with some
concrete new results, nor a review of a certain established
field.  Rather it is an attempt to formulate a problem to
single out a specific class of materials which have much in common and
which may have rather unique properties.  Thus this paper has in a
sense a conceptual character.  My aim is to attract attention to the
compounds with the negative charge transfer gaps, to discuss some of
their properties, to show the connection of this problem to some other
phenomena actively discussed nowadays such as insulator--metal
transitions, Kondo-insulators or High-$T_c$ Superconductivity.
The treatment is necessarily rather qualitative, and some of the
analogies I use may seem rather far-fetched.  But I hope that this
treatment may stimulate more active investigation of these compounds,
in the course of which some of the ideas put forth below may be either
confirmed or refuted.

For me personally, with my purely theoretical background, one of the
first expositions to the idea that the local properties of transition
metals {\sl at the atomic level} may be crucial for such phenomena as
insulator--metal transitions came from the discussions with R.~Dagys.
That is why I think that this article, in which these ideas play very
important role, can be appropriate for this issue.

\section{Transition-metal insulators}
As is well known, transition metal compounds (for concreteness we
shall speak below about oxides) have two main groups of electrons
determining their properties.  These are the electrons of partially
filled $d$-shells of TM, and $p$-electrons of
oxygen~\cite{goodenough}.  In the standard chemical language we
usually speak about certain valence, or oxidation state, of TM,
assuming that oxygen ions are O$^{2-}$. Thus in ionic picture we have
TM ions with partially filled $d$-shell ($d^n$) and oxygen ions with
filled $2p$-shell (configuration $p^6$).

The $d$-orbitals have relatively small radius and are rather
localized. In this case it is very important to take into account
Coulomb interaction of these electrons, in particular $d$--$d$
repulsion on the same site (Hubbard interaction)~\cite{hubbard}
$Un_{di\uparrow}n_{di\downarrow}$, where
$n_{di\sigma}=d^+_{i\sigma}d_{i\sigma}$,
$d^+_{i\sigma}$ are the creation operators of $d$-electrons on the
site $i$ with the spin $\sigma$ (we ignore for a moment orbital
index).  Just this interaction, when strong enough, makes typical
stoichiometric TM oxides insulators (Mott--Hubbard mechanism).

Another important group of electrons in these compounds are
$p$-electrons of oxygen. Coulomb interaction of these electrons can in
a first approximation be neglected (although sometimes it may be
important, especially in cases of large concentration of holes).
However one should definitely include the hybridization (covalent
mixing) of $p$-electrons with $d$-electrons of transition metals; this
hybridization determines many properties of these
materials~\cite{goodenough,anderson}.

An important parameter is the relative position of TM $d$- and
oxygen $p$-levels, $\epsilon_d$ and $\epsilon_p$, respectively. Thus
the simplest generic model describing TM compounds should have the
form
\begin{eqnarray}
H&\displaystyle=\sum\epsilon_dd^+_{i\sigma}d_{i\sigma}+\sum\epsilon_{p}p^+_{j\sigma}p_{j\sigma}+\sum\left(t_{pd}d^+_{i\sigma}p_{j\sigma}+{\rm
h.c.}\right)+
U\sum n_{di\uparrow}n_{di\downarrow}.\hfill\label{one}
\end{eqnarray}
Sometimes it is also necessary to take into account some extra
factors (orbital structure of $d$- and $p$-shells, anisotropy of
$p$--$d$ hopping, $p$--$p$ and $p$--$d$ interaction, direct $p$--$p$
hopping); however for the illustration the model~(\ref{one}) is
sufficient.

The model~(\ref{one}) can describe different situations:

{\it a.} \ If $p$--$d$ mixing (the third term in~(\ref{one}))
is small enough, the system (for one electron per site) is insulating.
Depending on the ratio of the parameters $U$ and
$\epsilon_d-\epsilon_p$ one may discriminate between two kinds of such
insulators~\cite{zaanen}.  If
\begin{equation}
\Delta=(\epsilon_d+U)-\epsilon_p>U,\label{two}
\end{equation}
an insulator is of a
Mott--Hubbard type: the lowest excited states corresponding to
the creation of charged excitations are the transitions of
$d$-electrons from one TM ion to another, $d^n+d^n\to
d^{n+1}+d^{n-1}$. To a first approximation then the energy gap for a
conductivity will be $E_g^{({\rm MH})}\sim U$.

Another situation exists if the oxygen $p$-levels are not deep enough.
Then the lowest charge excitations correspond to a transfer of an
electron from oxygen into TM $d$-shell, $p^6+d^n\to p^5+d^{n+1}$.
These excitations are lower than the Mott--Hubbard ones when
$\Delta<U$.  With our definition of parameters~(\ref{one}) the gap in
this case is $E_g^{({\rm Ch.tr.})}=\Delta$.  Such materials are
called charge-transfer insulators.  Corresponding phase diagram is
schematically illustrated in fig.~1.

Often one describes these
materials using not electron, but hole representation.  One can use in
this case the same Hamiltonian~(\ref{one}) but taking as a ground
state with zero energy the configuration $d^{n+1}p^6$ and treating
$d_{(\sigma)}^+p_{j\sigma}^+$ as hole creation operators, $\epsilon_d$
and $\epsilon_p$ as hole energy levels and $U$ as $d$--$d$ hole
repulsion.  Then the configuration $d^np^6$ corresponds to one
$d$-hole with the energy $\epsilon_d$, $d^{n-1}$---to two $d$-holes
with the energy $\epsilon_d+U$, and $p^5$---to a hole on oxygen with
the energy $\epsilon_p$.  Mott--Hubbard insulators then have a gap
$E_g=U$, and charge transfer insulators ---
$E_g=\epsilon_p-\epsilon_d=\Delta$; a borderline between them
again corresponds to $U\simeq\Delta$.

Most of the transition metal oxides belong to one of these two
categories.  They are insulators. Partially filled $d$-shells
determine their magnetic properties; an exchange interaction is
mostly due to a hopping via intermediate oxygen (superexchange
mechanism)~\cite{anderson}. Details of the resulting magnetic
structure depend on the type of $d$-orbitals occupied, on the
geometrical arrangement of corresponding ions etc.; it follows the so
called Goodenough--Kanamori--Anderson
rules~\cite{goodenough,anderson,kanamori}.

\section{Negative charge transfer gaps}
As it is clear from the previous discussion, there
may exist situations when oxygen $p$-levels lie high enough in
energy, so that the charge-transfer gap $\Delta$ may become very small
or even negative.  All the properties of corresponding systems could
change drastically.  One can encounter such a situation in
cases when the formal valence state of the transition metal is
unusually high.  In principle the valence of the TM can vary in wide
range, from 2+ up to sometimes 6+. Thus, there exist V oxides from VO
(V$^{2+}$) to V$_2$O$_5$ (V$^{5+}$); there are Cr compound ranging
from Cr$^{2+}$ (KCrF$_3$) to Cr$^{6+}$ (CrO$_3$, K$_2$CrO$_4$).

Usually materials at the beginning of the TM series (V, Ti, Cr), at
least for ordinary  valence, are believed to belong to the insulators
of Mott--Hubbard type.  With increase of valence they may cross to
charge-transfer regime (for higher valence the relevant
$d$-levels lie deeper). In principle they may also enter a regime with
very small $\Delta$ (we suspect that CrO$_2$ is such a case).

Heavier TM oxides usually belong to charge-transfer compounds. This is
now well established especially by high-energy spectroscopy (XAS,
XPS). It is shown e.g.\ that the holes
introduced by hole doping (for instance in
Ni$_{1-x}$Li$_x$O~\cite{kniper}) are mostly introduced into oxygen
$p$-states.  This fact plays also crucial role in High-$T_c$
Superconductors for which the basic structure elements are
CuO$_2$-planes~\cite{emery,zwezdin}.

One may argue that with the increasing valence of heavier
TM ions (Fe, Co, Ni, Cu) charge-transfer energy would decrease and may
become negative. In this case, instead of having an electronic
configuration, corresponding to the formal valence state, e.g.\
$d^np^6$, the system may prefer to have configuration $d^{n+1}p^5$.
One may say that TM ions ``resist'' having too high a valence and
prefer to retain lower valence (higher occupation of $d$-shell) at the
expense of creating holes on oxygens.

The resulting situation may be illustrated on a following scheme
(fig.~2). In fig.~2 we show the relative position of relevant
electronic states in different regimes. The case 2$a$) corresponds to
a Mott--Hubbard insulator: the chemical potential $\mu$ lies between
$d$-levels with different occupations and the lowest charge-carrying
excitations are those across the Mott--Hubbard gap, creating extra
electrons and holes in $d$-subbands ($d$-levels $\epsilon_d$ and
$\epsilon_d+U$ would become the lower and upper Hubbard bands). The
case~2$b$) corresponds to a charge-transfer insulator. Here lowest
excitations are those across the charge-transfer gap $\Delta$. We have
shown here also the split-off states (Zhang--Rice
singlets~\cite{zhang,eskes}) which can be created when the
hole is introduced into oxygen $p$-band (acceptor doping)
and which probably play crucial role in High-$T_c$
Superconductors.

The ``motion'' along the line AB in fig.~1 (decrease of $\Delta$)
corresponds here to the modification of an electronic structure shown
in figs.\ 2$c$,~2$d$.
We may have here the situation strongly resembling
the mixed-valent one, where we have partially filled oxygen band
overlapping and strongly hybridizing with the correlated $d$-states.

The notion of a negative charge-transfer energy (``negative
$\Delta$'') was introduced in a most apparent form by Fujimori et
al.~\cite{mizokawa} during the study of NaCuO$_2$. This material
which formally corresponds to Cu$^{3+}$ ($d^8$) was shown to have
rather the configuration $d^\alpha p^\beta$ with $\alpha\sim8.8$,
$\beta\sim5.2$---i.e.\ close to the configuration
Cu$^{2+}(d^9)$O$^-(p^5)$ (if one takes into account that there are two
oxygen per one copper, the average ``valence'' of oxygen will be
changed). There are other materials of the same class: probably to
them belong LaCuO$_3$; RNiO$_3$ (R---rare earth);
La$_{1-x}$Sr$_x$CoO$_3$; possibly the material with the ``colossal''
magnetoresistance La$_{1-x}$Ca$_x$MnO$_3$, etc.  Properties of these
compounds vary widely:  NaCuO$_2$ is a diamagnetic
insulator~\cite{hestermann,mizokawa}; LaCuO$_3$ is paramagnetic metal
(or small gap insulator---there is still some controversy
about it, see e.g.~\cite{czyzyk}).  LaNiO$_3$ is paramagnetic (Pauli)
metal, whereas the compounds RNiO$_3$ with $\rm R=Pr$, Nd, Sm have
insulator--metal transitions with the strange magnetic ordering in the
insulating phase~\cite{torrance,garcia}.  LaCoO$_3$ at low
temperatures is magnetic insulator with rather unusual magnetic
properties; it has gradual insulator--metal transition at high
temperatures~\cite{goodenough}.  However, doping by Sr (which formally
produces Co$^{4+}$) makes this system a ferromagnetic
metal~\cite{jonker,raccah}. The same is true for
La$_{1-x}$Ca$_x$MnO$_3$, see~\cite{goodenough,cro2}.  CrO$_2$, which
may also belong to this class~\cite{cro2}, is metallic and is one of
the few ferromagnetic TM oxides.  Thus we see that the compounds
belonging to this class have a wide variety of properties.

\section{Self-doping}
As is clear from figs.\ 1,~2, the situation with the small or
negative $\Delta$ should lead to drastic modification of all the
properties. In this case the $p$-band lies
very close or even higher than the initially empty $d$-levels so that
part of oxygen $p$-electrons should go to $d$-levels and there should
appear holes in $p$-band, fig~2$d$. One may call this situation {\it
self-doping\/}: here holes appear in $p$-band not due to externally
introduced impurities, but due to a special position of intrinsic
energy levels of the system.

Of course there always exists a hybridization between $p$-sites of
ligands and $d$-states of the transition metal ions, so that the
resulting bands have mixed character.  Thus one should not literally
treat the holes as belonging solely to oxygens.  Actually the wave
function describing the ground state of the system may be
schematically written as
\begin{equation}
\left|\Psi\right\rangle=\alpha\left|d\right\rangle+\beta\left|p\right\rangle
\end{equation}
and in case of negative charge-transfer gap the weight of the
configuration $|d^{n-1}p^5\rangle$ exceeds that of $|d^np^6\rangle$,
$|\beta|^2>|\alpha|^2$.  Keeping in mind the $d$--$p$ hybridization,
we can nevertheless say that in this case one should start not from
the configuration $|d^np^6\rangle$ corresponding to the formal
oxidation state of TM, but rather from the configuration
$|d^{n-1}p^5\rangle$ with the holes in $p$-band.  This extreme limit
is of course also an approximation, one should later on take into
account the hybridization and interaction of $p$-holes with the
remaining $d$-configurations, but for these compounds this state is
definitely a better starting point than the original
configuration~$d^np^6$.

This approach is now rather widely used in theoretical treatments of
the High-$T_c$ Superconductors, for which the $d$--$p$
model~(\ref{one}) (often called in this field the Emery model) is the
usual starting point, see below.  However the applicability of this
approach and even the necessity to use this model for the description
of wider class of TM compounds, especially those with small or
negative charge-transfer gap, is not yet widely appreciated, and many
phenomena connected with the corresponding physics are not
investigated well enough.

The situation with self-doping looks somewhat similar to that
in small- or zero-gap semiconductors (grey tin; HgTe) or semimetals
(Bi, graphite).  However there are also important differences. In
contrast to ordinary semimetals, here we have crossing of energy
levels of electrons with quite different character: one forming more
or less ordinary energy band (oxygen $p$-band), and another
intrinsically strongly correlated (localized $d$-electrons).
Consequently electron--electron interactions, both between the
$d$-electrons and between $d$- and $p$-electrons plays here very
important role.  In this respect the situation here resembles more
that in heavy-fermion or mixed-valence compounds, especially those
with small gaps (Kondo-insulators~\cite{aeppli}, like ``gold'' SmS,
SmB$_6$, CeNiSn etc). Relevant energy scales here, however, are
strongly different from the Kondo-systems, especially the $d$--$p$
hybridization is much stronger than in rare earth compounds. This
change of energy scale may in principle lead to different physical
behaviour, although an analogy of our problem with that in
valence-fluctuating rare earth compounds may be very productive.

\section{Properties of negative charge-transfer gap compounds}
The appearance of oxygen holes may lead to several important
consequences.  These holes are rather mobile: the effective bandwidth
of the $p$-band (actually, of course, hybridized $p$--$d$ band, which
however in this case has largely $p$-character) is
typically $\sim0.5\,\rm eV$.  Thus one could in principle expect that
the material would become a metal.

However this is not at all evident and need not in general be the
case.  Thus, in the undoped case one would have equal number of
$p$-holes and of extra electrons on $d$-shell. In this situation they
may prefer to form excitonic-like bound states, so that the resulting
state would still be an insulator (although with a gap of a collective
nature).  This is probably the case of NaCuO$_2$.  This state has much
in common with the above mentioned
Kondo-insulators~\cite{aeppli}---rare-earth compounds with equal
number of conduction electrons and localized spins, which also have
an energy gap of collective origin~\cite{khomskii2}.

In NaCuO$_2$ the
formation of this insulating ground state is probably facilitated by
the crystal structure of this compounds, which consists of
one-dimensional CuO$_2$-chains with the CuO$_4$-plaquettes sharing
common edge \cite{hestermann,mizokawa}, see fig.~3. However
other crystal structures---e.g.\ perovskite LaCuO$_3$ with
three-dimensional network of CuO$_6$-octahedra sharing common
cor\-ners\hskip0pt---may be more favourable for a metallic state
\cite{mizokawa} (although it is not really proven, see the discussion
in~\cite{czyzyk}).

If the resulting ground state would be a metal, it may be rather
unusual: it will be a metal of a mixed-valence type, where one of
the components---$d$-orbitals admixed into a conduction band---is
strongly correlated whereas another one---oxygen $p$-electrons---may
be treated as uncorrelated.  What is the nature of the corresponding
insulator-metal transition due to closing of the charge-transfer
gap, is a completely open problem.

One can have here in principle two possibilities, with even or odd
number of electrons per elementary cell.  They are illustrated in
figs.~4,~5 (cf.\ fig.~2); for simplicity we speak here about
electrons.

In fig.~4 we show possible evolution of spectra when $d$-level with
even number (e.g.\ two) $d$-electrons moves towards empty conduction
band\footnote{For oxides this would correspond to a hole
representation: empty $d$-level approaching full conduction band, see
fig.~2.}.
We see that here we have insulators both for $\epsilon_c\gg\epsilon_d$
(for oxides it would mean $\epsilon_p^{\it
hole}\gg\epsilon_d^{\it hole}$) and in the opposite limit
$\epsilon_d\gg\epsilon_c$, where both electrons are in a conduction
band.  In this case we have good chances to have insulator also in
between, which would correspond to a Kondo-insulator (one $d$-electron
per one electron in a conduction band)---although it is not proven and
may in principle depend on the specific details such as crystal
structure etc.  The situation illustrated in fig.~4 coresponds to a
low-spin state.  If the initial configuration would correspond to a
high-spin state (parallel spins in fig.~$4a)$) there should occur
magnetic--nonmagnetic transition somewhere in between.

Fig.~5 corresponds to an odd number of electrons per unit cell. Here
fig.~5$a$ (deep $d$-level) describes the (magnetic) insulator of a
Mott-Hubbard type.  However an opposite limit
($\epsilon_d\gg\epsilon_c$) would correspond to a metal with all the
electrons in a partially filled conduction band and an empty
$d$-level.  Thus we should have somewhere in between (when
$\Delta=\epsilon_c-\epsilon_d$ changes sign and becomes negative) an
insulator--metal transition. There is at present no theory of such
transitions, e.g.\ it is not clear whether they would be I-st or II-nd
order, whether magnetic order disappears simultaneously with
the insulator--metal transition, etc.

\looseness=1
The situation with the negative charge-transfer gap may lead to a
specific modification of the local electronic configuration
(crystal-level scheme) of the TM ion itself.  In particular the
configurations which are usually unstable may be stabilized in this
case~\cite{potze-saw}.  Let us consider this situation on the specific
example of the compounds containing Co$^{3+}(d^6)$, e.g.\
LaCoO$_3$~\cite{korotin}.  Usually two competing configurations are
considered for this case: the high-spin ($S=2$) and the low-spin
($S=0$) ones, see fig.~6~$a),b)$.  If we now treat the situation with
the negative charge-transfer gap, we should have locally the
configuration which is rather $d^7p^5$.  The ion with the
configuration $d^7$ should definitely be in the high-spin state with
$S_d=\frac32$, see fig.~$6c)$.  However there will be an oxygen hole
nearby which is schematically illustrated in fig.~6$c)$.  The
hybridization of oxygen $p$-electrons with the predominantly
$e_g$-states of the TM is rather strong, and it always decreases the
total energy if corresponding mixing is allowed.  From fig.~6$c)$ it
is clear that for that the spin on oxygen should be antiparallel to
the spin of the TM (here Co).  Thus the total spin of this state will
be $S_{\it tot}=\frac32-\frac12=1$, i.e.\ the configuration Co$^{3+}$O
($d^7p^5$) will have neither high-, nor low-spin state, but an
intermediate spin $S=1$.  Recent band-structure calculation using
LDA+U method which permits to take into account electron correlations,
indeed confirmed that an intermediate spin state may be the ground
state of LaCoO$_3$ at certain conditions~\cite{korotin}.  This gives
an alternative explanation of the intriguing properties of
LaCoO$_3$~\cite{goodenough} which were usually interpreted in a
framework of the model of the low-spin--high-spin transition.

Another interesting group of problems emerge when we think of the
possible magnetic structure of these materials. As is mentioned above,
in ordinary cases (Mott--Hubbard or charge-transfer insulators with
filled oxygen $p$-shells) we can describe magnetic interactions as
mostly due to superexchange. The rules which determine magnetic
structure (Goodenough--Kanamori--Anderson
rules~\cite{goodenough,anderson}) are well known and are very
successful.

In case of negative charge-transfer energy the situation may be
modified drastically. If we have, besides $d$-electrons ($d$-spins),
also holes (spins) on oxygen, then one may expect different exchange
mechanisms entering the game.  If oxygen holes are delocalized and
form bands, there may appear $d$--$d$ interaction of the RKKY-type via
$p$-band. Or, if a band is sufficiently narrow, double-exchange
mechanism~\cite{zener} (tendency to form ferromagnetic alignment of
localized spins provided by mobile holes) may be efficient; it may
produce e.g.\ ferromagnetic metallic state (this may be the
situation in CrO$_2$, La$_{1-x}$Ca$_x$MnO$_3$ and in
La$_{1-x}$Sr$_x$CoO$_3$).  If, on the other side, oxygen holes could
be treated as localized, there should be strong $p$--$d$ exchange
interaction which would also give a ferro- (or rather ferri-) magnetic
state: the spins of TM ions will be parallel, and antiparallel to
the spins of oxygens sitting in between~\cite{aharoni},
see fig.~7\ \footnote{We consider here the crystal structure like that
in perovskites or in La$_2$CuO$_4$; in principle the outcome should be
sensitive to the detailed crystal structure.}.

The role of orbital degrees of freedom in negative-$\Delta$ systems is
also far from trivial. It is known in particular that in case of
orbital degeneracy the system becomes unstable (Jahn-Teller effect),
and in concentrated case it experiences the cooperative
Jahn-Teller transition~\cite{kugel2} (an orbital ordering accompanied
by the structural phase transition; this is one of the very few cases
of structural transitions for which we know its microscopic origin).
What would become of these transitions for negative $\Delta$, is not
known.  The change of the occupation of $d$-level may remove the
degeneracy, or, vice versa, may make degenerate the system which by
formal valence is not degenerate.  However, the degeneracy removed
from $d$-orbitals may still be present (it may be ``transferred'' to
the degeneracy of the corresponding oxygen holes).  What will be the
outcome is definitely interesting to explore; this may determine many
properties of these compounds, both structural and magnetic, and may
be crucial, e.g.\ for the ``colossal'' magnetoresistance in
La$_{1-x}$Ca$_x$MnO$_3$.

There is also an important connection of the situation in
nega\-tive-$\Delta$ compounds and in doped cuprates (High-$T_c$
Superconductors).
In these latter materials the holes introduced by doping form singlet
bound states with Cu$^{2+}$ spins (Zhang-Rice
singlets~\cite{zhang,eskes}) which most probably determine transport
and superconducting properties at relatively low doping. In this sense
the substances with the formal valence Cu$^{3+}$ (NaCuO$_2$;
LaCuO$_3$) could correspond to a situation where we have these
singlets {\it at each site}, cf.\ fig.~3. However the very picture of
singlet bound states may turn out to be invalid for very high
concentration of them: excessive doping may lead not to a {\it
``removal''} of spins from our system (which is the case with
Zhang-Rice singlets and which is usually described by $t$--$J$
model~\cite{zhang}) but to {\it adding} extra spins (we have all
$d$-electron spins and we add holes---or spins---on
oxygen~\cite{emery2}). What will be the properties of such systems is
not really clear. We thus see that our problem (description of
nega\-tive-$\Delta$ compounds) has much in common with the description
of overdoped High-$T_c$ Superconductors.

In this connection one more remark is in order.  Until now we
discussed mostly the stoichiometric compounds with negative
charge-transfer gap.  However most probably it is just in doped
materials that the specific properties of these systems could be most
pronounced.  Indeed, as we mentioned above, $p$--$d$ hybridization is
always present in the TM compounds, and as a result the electronic
configuration is never really $d^np^6$ but always has admixture of
other configurations, such as $d^{n+1}p^5$.  In that sense the
difference between the usual compounds and those with negative CTG may
be sometimes more quantitative than qualitative, according to
Eq.~(\ref{two}) and to the discussion there.  It is true that the
theoretical discussion in this case should start from the opposite
limit of holes in the $p$-band, but the outcome may in some cases look
similar to the usual treatment.  However, for doped compounds we have
more chances to have really different behaviour.  The case of
High-$T_c$ cuprates gives a good example of the new striking
phenomena which may be connected with the introduction of holes into
oxygen $p$-states.  One can even speculate: does not the clue to the
search of new (high-temperature) superconductors lie in the unusually
high valence?

\section{Conclusion}
We see that the TM compounds with the unusually high valence of
the TM ions and small or negative charge-transfer gap constitute
well-defined special class of compounds, for which we can expect
rather wide range of properties.

Experimental studies of these compounds are now under way. However
their theoretical understanding is still lacking.
The nature of the resulting states is a completely open problem: at
the moment we do not know when to expect metallic or insulating
behaviour (or the insulator-metal transition), what will be
the magnetic properties, etc. There are clear links of this problem to
that of heavily doped High-$T_c$ Superconductors and to the
Kondo-insulators, and one may hope that eventual progress in the
description of negative-$\Delta$ systems may shed some light on these
problems too (and vice versa).

Experimentally there are also a lot of problems to be solved.   One
has to check whether all the ``suspicious'' materials mentioned above
(LaCuO$_3$; La$_{1-x}$Sr$_x$CoO$_3$; PrNiO$_3$; CrO$_2$;
La$_{1-x}$Sr$_x$MnO$_3$) belong to the class of compounds with small
or negative charge-transfer gap.  Probably the most direct tool for
this is high-energy spectroscopy and possibly some local spectroscopic
probes (NMR, ESR).  Special attention should be paid to the properties
of doped compounds of this type.  Thus for example it would be very
interesting to try to dope, e.g.\ by Sr or Ba, the
quasi-onedimensional compounds NaCuO$_2$ or KCuO$_2$, which
definitely have negative charge-transfer gap (``Zhang--Rice singlets
at each site'').  This would be the counterpart of the High-$T_c$
cuprates in which by doping e.g. La$_2$CuO$_4$ we go formally from
Cu$^{2+}$ in the direction of ``Cu$^{3+}$'': here it would correspond
to moving in the opposite direction, from ``Cu$^{3+}$'' toward
Cu$^{2+}$.  What would be the properties of such system would be very
interesting to study.

One can raise many other interesting questions and suggest a lot of
other experiments to probe the behaviour of these systems.  Here I
only want to stress once again that the study of this very specific
and interesting class of transition metal compounds presents definite
interest both from the fundamental point of view and possibly for
certain practical applications.

\bigskip
\noindent\hfil{\bf Acknowledgements:} \ I am very grateful to
G.~A.~Sawatzky and D.~van~der~Marel for the useful discussion.  This
work was supported by the Netherlands Foundation for Fundamental Study
of Matter (FOM).

\newpage
\def\B #1/#2/#3/#4/#5/{#1, #2 {\bf #3}, #4 (#5)}

\bigskip

\noindent P.S. (December 2000)\hfil\break
This small, review-like, paper was written in 1997 and published in
the memorial issue of the Lithuanian Journal of Physics {\bf 37}, 65
(1997) devoted to the memory of my late friend R.~Dagys.  As this
journal is not widely available, I think it worthwhile to put it on
the web, although it is already three years old.  The development
since 1997 has shown that the problems  raised in this paper are
still important, and the discussion presented in this paper may still
be useful.  I did not change the main content of the paper; to
somewhat update it, below are two more recent references relevant to
the problems discussed in this paper.

\begin{description}

\item \B M.~A.~Korotin, V.~A.~Anisimov, D.~I.~Khomskii and
G.~A.~Sawatzky/Phys.~Rev.~Lett./80/4305/1997/

\item \B T.~Mizokawa, D.~I.~Khomskii and
G.~A.~Sawatzky/Phys.~Rev.~B/61/11263/2000/

\end{description}

\newpage
\section*{Figure Captions}
\begin{description}
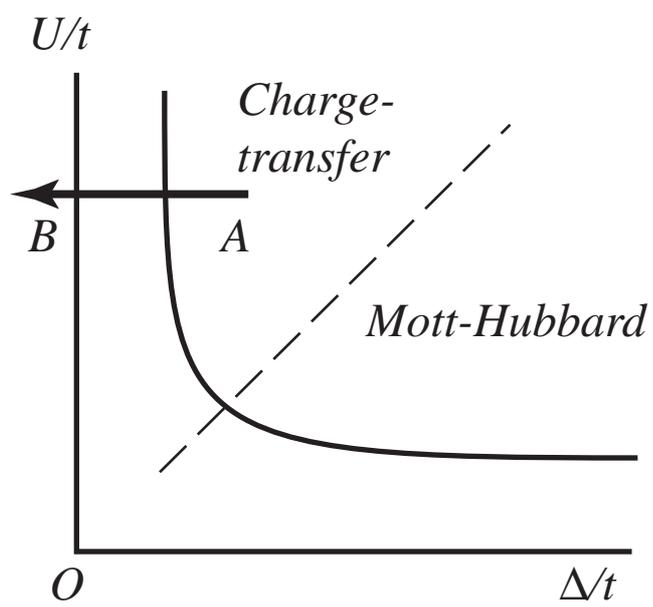
\item{Fig.~1 \ }Schematic phase diagram of the transition metal
compounds (by~\cite{zaanen})

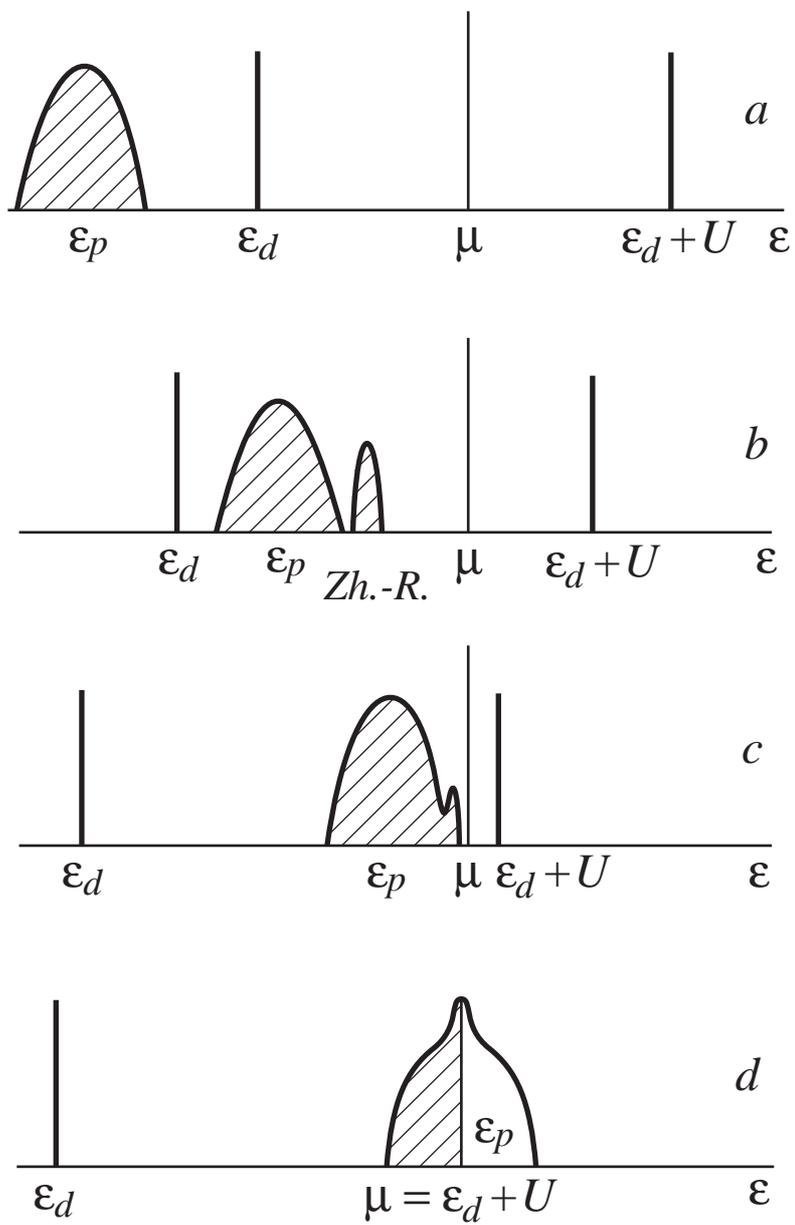
\item{Fig.~2 \ }Energy spectrum of \ $a)$~Mott-Hubbard insulator; \
$b)$~charge-transfer insulator with the positive charge-transfer gap
$\Delta$; \ $c),d)$~the system with small and negative charge-transfer
gap

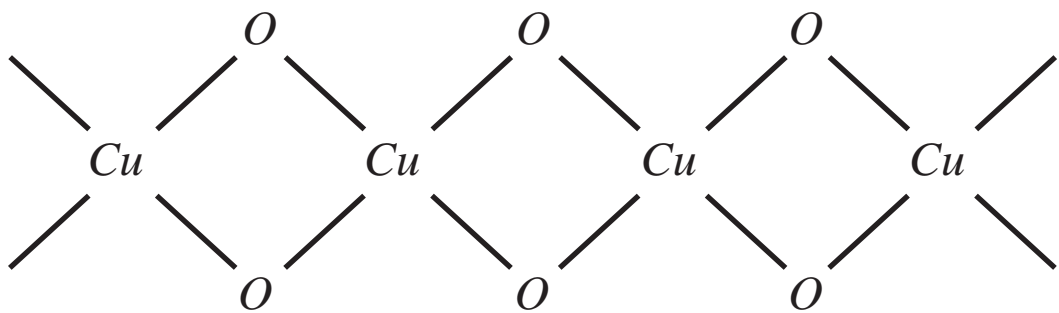
\item{Fig.~3 \ }The main motive of the crystal structure of NaCuO$_2$

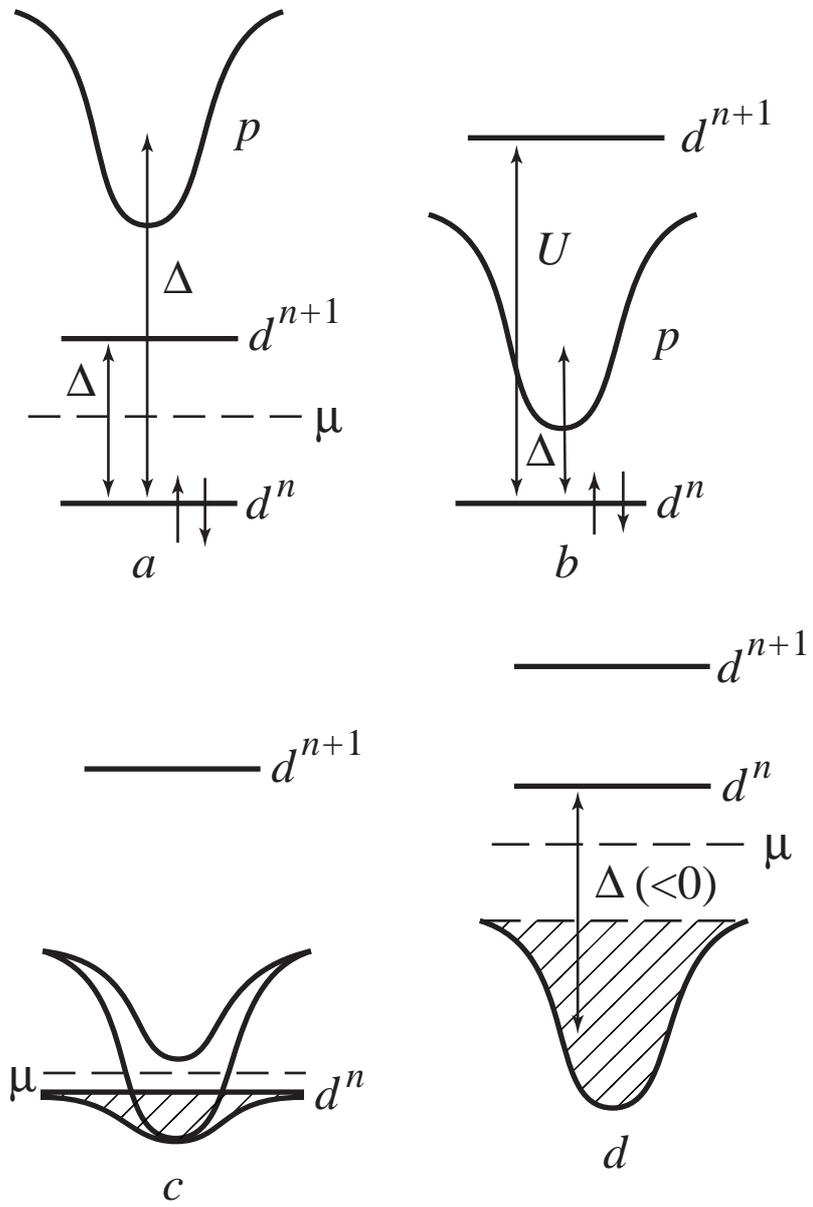
\item{Fig.~4 \ }Evolution of the energy spectrum with the even
(e.g.\ two) number of electrons per unit cell \ $a)$~Mott-Hubbard
insulator (low-spin state); \ $b)$~charge-transfer insulator
($\Delta>0$); \ $c)$~charge-transfer insulator ($\Delta\sim0$); \
$d)$~the system with the negative charge-transfer gap; in this
particular case it is also an insulator.  $\mu$ is the chemical
potential

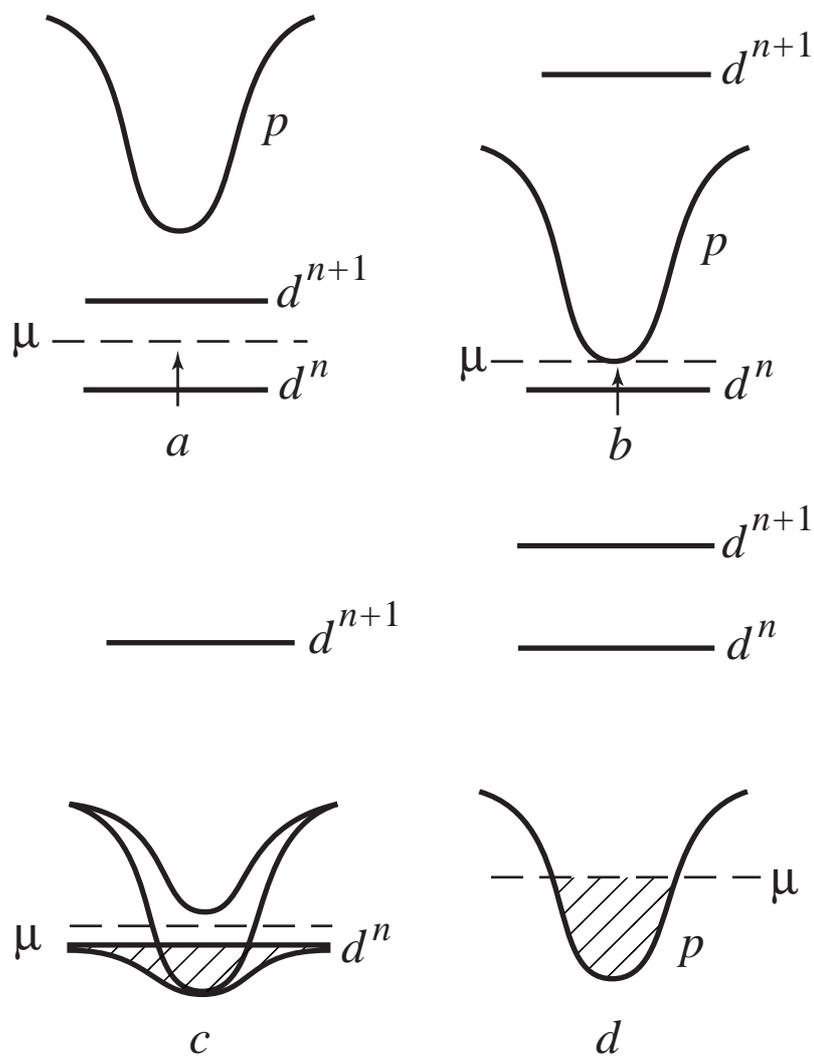
\item{Fig.~5 \ }Evolution of the energy spectrum with the odd
number (here one) of electrons per unit cell \ $a)$~Magnetic
Mott--Hubbard insulator; \ $b)$~magnetic charge-transfer insulator; \
$c)$~the situation with the small charge-transfer gap, which can be an
insulator or a metal; \ $d)$~the situation with large negative
charge-transfer gap, which in this particular case should be a metal
with empty $d$-shell.  Other notations are the same as in fig.~4

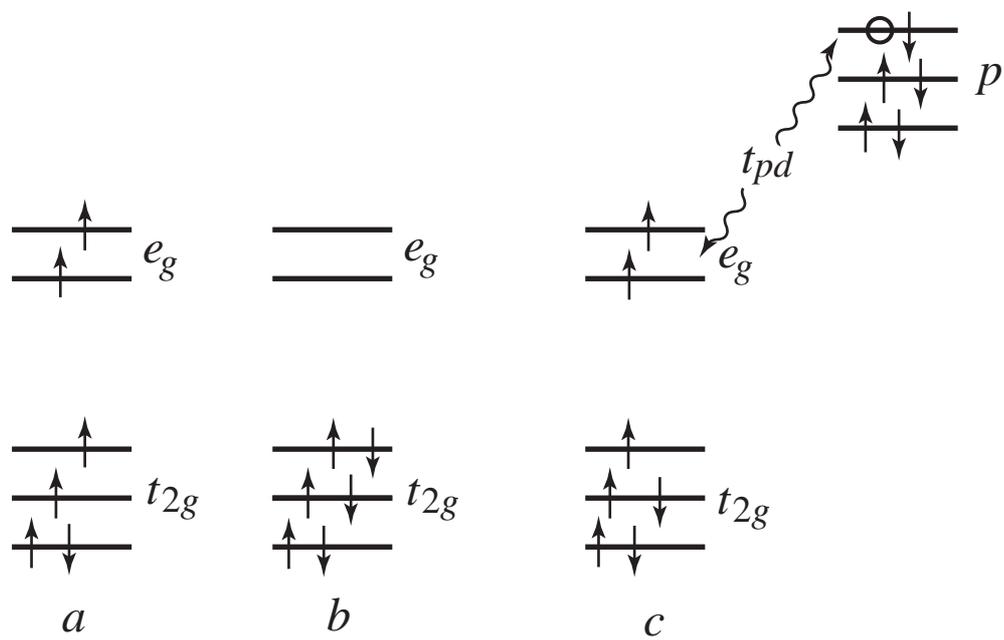
\item{Fig.~6 \ }Energy levels and electron occupation of the ion with
the configuration $d^6$ (e.g.\ Co$^{3+}$): \ $a)$~high-spin state; \
$b)$~low-spin state; \ $c)$~intermedi\-ate-spin state.  $t_{pd}$ is
the transition metal--oxygen hybridization

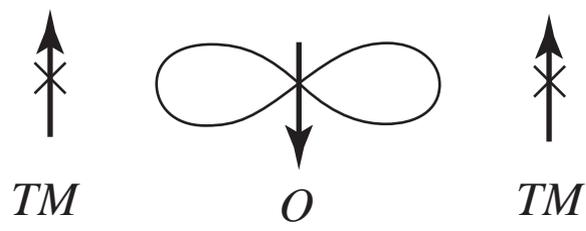
\item{Fig.~7 \ }Possible origin of the ferromagnetic ordering in case
of localized oxygen $p$-holes (by~\cite{aharoni})
\end{description}

\end{document}